\documentclass[twocolumn,showpacs,preprintnumbers,amsmath,amssymb,nofootinbib,pra,superscriptaddress,10pt]{revtex4-1}

\usepackage{graphicx}
\usepackage{dcolumn}
\usepackage{bm}
\usepackage{epstopdf}

\usepackage{amssymb} 
\usepackage{amsmath} 
\usepackage{latexsym} 
\usepackage{amsthm} 
\usepackage{mathrsfs} 

\theoremstyle{plain}

\theoremstyle{definition}

\theoremstyle{remark}

\begin{document}
\def \beq{\begin{equation}}
\def \eeq{\end{equation}}


\title{Metriplectic Structure of a Radiation-Matter Interaction Toy Model}

   \author{Giulia Marcucci}
   \affiliation{Department of Physics, University Sapienza, Piazzale Aldo Moro 5, 00185 Rome (IT).}
   \affiliation{Institute for Complex Sy\-Stems, National Research Council (ISC-CNR), Via dei Taurini 19, 00185 Rome (IT).}
    \email{giulia.marcucci@uniroma1.it}

    \author{Claudio Conti}
   \affiliation{Institute for Complex Sy\-Stems, National Research Council (ISC-CNR), Via dei Taurini 19, 00185 Rome (IT).}
\affiliation{Department of Physics, University Sapienza, Piazzale Aldo Moro 5, 00185 Rome (IT).}

\author{Massimo Materassi}
   \affiliation{Institute for Complex Sy\-Stems, National Research Council (ISC-CNR), Via Madonna del Piano 10, 50019 Sesto Fiorentino, Florence (IT).}
      \affiliation{National Institute of Astrophysics, Rome - Institute for Space Astrophysics and Planetology (INAF-IAPS).}

\date{\today}

\begin{abstract}

A dy\-na\-mi\-cal sy\-stem defined by a metriplectic structure is a dissipative model characterized by a specific pair of tensors, which defines the Leibniz brackets. Generally, these tensors are Poisson brackets tensor and a symmetric metric tensor that models purely dissipative dynamics.

In this paper, the metriplectic sy\-stem describing a simplified two-photon absorption by a two-level atom is disclosed. The Hamiltonian component describes the free electromagnetic ra\-dia\-tion. The metric component encodes the ra\-dia\-tion-matter coupling, driving the sy\-stem to an asymptotically stable state in which the excited level of the atom is populated due to absorption.

This work is intended as a first result to pave the way to apply the metriplectic formalism to many other irreversible processes in nonlinear optics.

\end{abstract}

\maketitle

\section{\label{sec:intro}Introduction}
The modeling of irreversible sy\-stems is a fundamental issue in each physical field. Even though in quantum mechanics is a still debated topic, and many efforts have been done both in case of intrinsic irreversibility~\cite{Marcucci2016} and in open sy\-stems~\cite{Garmon2012}, classical mechanics boasts many more tools and much more established and recognized theories to describe time asymmetric phenomena. Nevertheless, when considering irreversibility due to dissipation, the study of metriplectic structures unveils a simple theory based on linear algebraic tools that have immediate thermody\-na\-mi\-cal translation, both in classical~\cite{Turski1987} and in quantum~\cite{Turski1996} sy\-stems. 

In li\-te\-ra\-ture there are several examples of irreversible dynamics represented as metriplectic sy\-stems: from very simple sy\-stems in Newton's mechanics~\cite{Materassi2012bis}, to hydrodynamics~\cite{Morrison1984} and magneto-hydrodynamics~\cite{Materassi2012}; more delicate, but extremely interesting, are the cases of kinetic equations, the collisional terms of which may be written as a semi-metric term, or that of a free rotator driven to a stable rotation axis by a suitably designed servo-engine~\cite{Morrison1986}. In all the cases mentioned the non-Hamiltonian sy\-stem gets gifted of the transparency of motions generated by Leibniz algebrae, even if proper symplectic formalism is not applicable; moreover, the e\-ner\-gy landscape becomes tractable as the free e\-ner\-gy is explicitly written.

The process of two-photon absorbtion (TPA) by a two-level atom is here described through a \textit{classical} dy\-na\-mi\-cal sy\-stem, in which the e\-ner\-gy initially located in the ra\-dia\-tion va\-ria\-bles is \textit{irreversibly} converted into the e\-ner\-gy pertaining to the po\-pu\-la\-tion of the excited level. The final state, in which no free ra\-dia\-tion exists any more while the excited state is populated, is \textit{the asymptotic e\-qui\-li\-brium state} of the sy\-stem. The existence of asymptotically stable e\-qui\-li\-brium makes the TPA similar to a \textit{dissipative process}, like macroscopic friction, where an ``ordered'' form of e\-ner\-gy is ``consumed'' in favour of the ``internal e\-ner\-gy'' of some medium. This attitude describes the matter absorbing the electromagnetic wave e\-ner\-gy as the environment of a sy\-stem that would be Hamiltonian \textit{per se}: the presence of the environment, with the matter-ra\-dia\-tion interaction that ``destroys'' the ra\-dia\-tion, breaks the Hamiltonian nature of the ra\-dia\-tion dynamics. Such a scenario is described by the extension of the symplectic algebra of the Hamiltonian sy\-stem to a \textit{metriplectic algebra of brackets}~\cite{Morrison1986}, where the Hamiltonian component of the motion is still given by the original Poisson bracket, while a suitable \textit{semi-defined metric bracket} generates the non-Hamiltonian component. An extension of the Hamiltonian, namely the \textit{free e\-ner\-gy} of the sy\-stem, represents the metriplectic generator of the motion. The foregoing program interprets the dynamics of classical dissipative sy\-stems as flows generated by a new kind of \textit{Leibniz algebrae} of brackets~\cite{Guha2007}, namely the \textit{metriplectic bracket}.

This paper is organized as follow.
In Section~\ref{sec:prefirst} we review the metriplectic formalism from a very general point of view.
In Section~\ref{sec:first} the dy\-na\-mi\-cal va\-ria\-bles are presented, together with the ODEs describing their evoulution in the presence of the dissipative interaction. Then, the dissipationless, i.e. Hamiltonian, limit is presented, in which the expression of the free ra\-dia\-tion e\-ner\-gy $H_0 (q,p)$ works as a Hamiltonian and the po\-pu\-la\-tion $n$ does not evolve.
In Section~\ref{sec:second} the metriplectic algebra generating the non-Hamiltonian component of the dynamics is constructed: first of all, equations are composed to define the semi-metric tensor through which the metric bracket $\left( \cdot,\cdot \right)$ is defined; then, a completion e\-ner\-gy $U(n)$ is constructed in order for $H_0 (q,p) + U(n)$ to be constant along the non-Hamiltonian motion of the full sy\-stem $(q,p,n)$. Finally, the framework is completed by defining the proper conditions on the entropy $S(n)$ and writing down the expression of the free e\-ner\-gy $F(q,p,n) = H(q,p) +\chi S(n)$, being $H=H_0+U$ and the e\-qui\-li\-brium points are determined as a consequence of this construction (in the sense that, choosing different expressions for $S(n)$, i.e. for $F(q,p,n)$, different e\-qui\-li\-bria $n_{\mathrm{eq}}$ are found).
In Section~\ref{sec:conclusions} we summarize the results of our analysis.
Details on the computation of the metric tensor $G$ are added in Appendix.

\section{\label{sec:prefirst} General Metriplectic Formalism}

Before describing how the metriplectic formalism is applied to the TPA, it is useful to sketch briefly the construction of a metriplectic sy\-stem. Typically, one starts from a Hamiltonian sy\-stem described by a set of va\-ria\-bles $X$, the dynamics of which is generated by some Hamiltonian $H_0(X)$ and some Poisson bracket $\left\{ \cdot,\cdot\right\}$ so that $\left( \dot{X} \right)_0 = \left\{ X,H_0(X) \right\}$ (the subscripts ``0'' refer to the dynamics generated by the sole $H_0$ via the bracket $\left\{ \cdot,\cdot\right\}$). Then, some quantity $S$ is defined, with the pro\-per\-ty of being \textit{in involution} with \textit{any} possible function of $X$, $\left\{ S,A\right\} = 0 \ \forall \ A(X)$, i.e. to be \textit{a Casimir of $\left\{ \cdot,\cdot\right\}$}: this quantity $S$ may either depend on the original va\-ria\-bles $X$ only (as for the kinetic theories or for the rigid body), or on some ``environmental'' variable $Y$ too (as in the case of a particle motion with friction, or those of non-ideal hydrodynamics or magneto-hydrodynamics: this will be the case here too). The Casimir $S$ becomes the generator of a new non-Hamiltonian component of the motion, through the introduction of a new bracket, $\left(\cdot,\cdot\right)$, with properties of symmetry and semi-definiteness~\cite{Morrison1986}: the extended sy\-stem, based on the old Hamiltonian one, has now a new dynamics in which the va\-ria\-bles $X$ evolve according to
\begin{equation}
\dot{X}=\left\{ X,H\left(X,Y\right)\right\} +\chi\left(X,S\left(X,Y\right)\right),\label{eq:metriplectic.motion.X}
\end{equation}
while the motion of the environmental va\-ria\-bles, if any, is typically influenced by $S$ and the metric bracket only:
\begin{equation}
\dot{Y}=\chi\left(Y,S\left(X,Y\right)\right).\label{eq:metriplectic.motion.Y}
\end{equation}

In Eq.~(\ref{eq:metriplectic.motion.X}) the Hamiltonian $H(X,Y)$ may be different from the original $H_0(X)$, as it may include a term depending on $Y$ in order to close the sy\-stem energetically, and take into account of the irreversible consumption of $H_0(X)$ (dissipation): the difference $U=H-H_0$ is the \textit{internal e\-ner\-gy} of the environment. In  Eqs.~(\ref{eq:metriplectic.motion.X}, \ref{eq:metriplectic.motion.Y}), the factor $\chi$ is a coefficient representing a coupling condition between $X$ and $Y$, or characterizing the asymptotically stable e\-qui\-li\-brium; the strength of the dissipative interaction, defining the non-dissipative (Hamiltonian) regime in some suitable limit of its, is some $\alpha$ included in the de\-fi\-ni\-tion of $\left(\cdot,\cdot\right)$, so that $\alpha\rightarrow0$ turns off dissipation. One may well say:
\begin{equation}
\lim_{\alpha\rightarrow0}\dot{X}=\left\{ X,H_{0}\left(X,Y\right)\right\} ,\ \ \lim_{\alpha\rightarrow0}\dot{Y}=0.\label{eq:kappa.tends.to.0}
\end{equation}
(from Eq.~(\ref{eq:metriplectic.motion.X}, \ref{eq:metriplectic.motion.Y}) it appears that also the limit for $\chi\rightarrow 0$ gives the ODEs in Eq.~(\ref{eq:kappa.tends.to.0}); however, this does not switch off dissipation, but simply describes a condition in which it is uneffective, see Sections~\ref{sec:second} and~\ref{sec:conclusions}).

As far as the bracket $\left( \cdot,\cdot \right)$ and the Casimir $S$ are concerned, the semi-definiteness of the first one, $\left(A,B\right)\le0\ \forall\ A,B$, and a suitable choice of the sign of $\chi$, i.e. $\chi < 0$, implies that $S$ will grow monotonically along the sy\-stem motion $\dot{S}\ge0$, until some asymptotically stable e\-qui\-li\-brium $Z_{\mathrm{eq}}=(X_{\mathrm{eq}},Y_{\mathrm{eq}})$ is reached, so that $\dot{S}(Z_{\mathrm{eq}})=0$~\cite{Morrison1986}. In other words, the Casimir $S$ turns out to be \textit{a Lyapunov function around $Z_{\mathrm{eq}}$}, and it can be understood as a form of \textit{entropy}~\cite{Materassi2016} (of course, all the reasoning just presented is rephrased ``without $Y$'' for those metriplectic sy\-stems in which no ``environment'' needs to be defined).

In order to complete the picture, the pro\-per\-ty $\left(H,A\right)=0\ \forall\ A$ is requested for the metric bracket and the total Hamiltonian $H$, in order for dissipation not to ``delete'' the total e\-ner\-gy, but just transform it. It must be underlined that this construction does not include ``all'' the dy\-na\-mi\-cal sy\-stems referred to as ``metriplectic'' in li\-te\-ra\-ture: this is the construction of a complete metriplectic sy\-stem (CMS), while incomplete metriplectic sy\-stems (IMS) may be defined too, with the two brackets but the Hamiltonian as the only generator. IMS are suitable to describe energetically open sy\-stems~\cite{Turski1987}.

The development presented here makes the TPA process tractable in a very transparent way as a CMS, and points towards the sy\-stematic algebrization of non-linear optics.

The sy\-stem introduced here in order to turn the TPA process into a CMS has three degrees of freedom: ra\-dia\-tion is represented either via a complex phasor $\psi$, or a couple of real va\-ria\-bles $(q,p)$; the po\-pu\-la\-tion of the excited level is given by some real variable $n$. During the \textit{irreversible process}, the electromagnetic e\-ner\-gy $H_{0}(q,p)$ is converted into the e\-ner\-gy $U(n)$ associated with $n\neq0$. In our ``metriplectization'' scheme one starts from the equations of motion of the state $Z=(q,p,n)$ and observes that a suitable limit of them reduces the sy\-stem to a Hamiltonian one. In this Hamiltonian limit a Poisson bracket is defined, so that $q$ and $p$ are canonically conjugated $\left\{ q,p\right\} =1$, while $n$ remains apparently outside the play as $\left\{ n,q\right\} =\left\{ n,p\right\} =0$.
Indeed, the po\-pu\-la\-tion of the excited level is \emph{in involution} with $q$ and $p$, so that \textit{any} function $S(n)$ will be a Casimir for $\left\{ \cdot,\cdot\right\} $.
The program then is to find a suitable function $H_0(q,p)$ that may play the role of Hamiltonian in the Hamiltonian limit: this represents the free ra\-dia\-tion e\-ner\-gy, to be extended as $H(q,p,n)=H_0(q,p)+U(n)$ to include the e\-ner\-gy pertaining to the filling of the excited state, namely the internal e\-ner\-gy of the environment ``atoms''. In order to complete the metriplectic framework, suitable forms for $S(n)$ and for the metric bracket $\left( \cdot,\cdot \right)$ must be constructed, and this is essentially what is done in the present work.

\section{\label{sec:first} Two-Photon Absorption Toy Model}
\begin{figure}[h]
\centering
\includegraphics[width=\columnwidth]{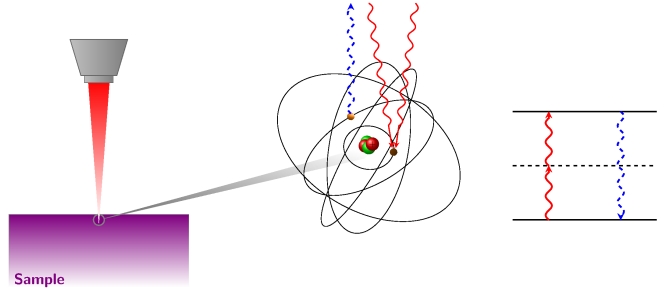}
\caption{Pictorial sketch of absorbtion of two photons in a two-level atom. In our sy\-stem, Eq.~(\ref{eq:2levels}) does not have terms of spontaneous emission, considered ne\-gli\-gi\-ble. This is here represented by the dashed blue line, not present in our model.}
\label{fig:TPA}
\end{figure}

We consider a very simplified toy model for the two-level atomic sy\-stem~\cite{Boyd2008,Moloney2004}. The TPA, sketched in Fig.~(\ref{fig:TPA}), is expressed by the following differential equations:
\beq
\left\{\begin{array}{l}\dot{\psi}=\imath|\psi|^2\psi-\alpha n \psi \\ \dot{n}=\frac k2|\psi|^4\end{array}\right. ,
\label{eq:2levels}
\eeq
being $\psi$ the complex field amplitude, $n$ the po\-pu\-la\-tion of second level, $k>0$ and $\alpha>0$.
This sy\-stem is directly derived by the multi-photon absorption model~\cite{Feng1997, Mlejnek1998, Schwarz2000}, when neglecting several physical phenomena, e.g., the spontaneous emission.
Moving to real-valued functions, we define
\beq
\psi=\frac{q-\imath p}{\sqrt{2}},
\eeq
so that in terms of the va\-ria\-bles $q$ and $p$ the sy\-stem~(\ref{eq:2levels}) reads:
\beq
\left\{\begin{array}{l}\dot{q}=\frac 12 p(q^2+p^2)-\alpha n q\\ \dot{p}=-\frac 12 q(q^2+p^2)-\alpha n p\\ \dot{n}=\frac k8(q^2+p^2)^2 \end{array}\right. .
\label{eq:2lev_qp}
\eeq
It is useful to observe that, in the limit
\beq
\alpha\rightarrow 0, \;  k\rightarrow 0 
\label{eq:nndiss}
\eeq
Eqs.~(\ref{eq:2lev_qp}) become a Hamiltonian sy\-stem, so that the conditions~(\ref{eq:nndiss}) will be referred to as \emph{non-dissipative limit} (NDL); under these conditions, the ODEs in $q$ and $p$ read:
\begin{equation}
\left\{ \begin{array}{l}
\dot{q}=\frac{1}{2}p(q^{2}+p^{2})\\
\dot{p}=-\frac{1}{2}q(q^{2}+p^{2})\\
\dot{n}=0
\end{array}\right. .
\label{eq:k0alpha0ODEqp}
\end{equation}
As one defines the Hamiltonian
%
\beq
H_0=\frac12\left(\frac{q^2+p^2}2\right)^2.
\label{eq:h0_qp}
\eeq
and the Poisson bracket
\begin{equation}
\left\{ q,p\right\} =1,\ \left\{ q,n\right\} =0,\ \left\{ p,n\right\} =0,\label{eq:Poisson.bra.1}
\end{equation}
any quantity $f\left(q,p,n\right)$ evolves according to:
\[
\dot{f}=\left\{ f,H_{0}\right\} 
\]
along the motion~(\ref{eq:k0alpha0ODEqp}). The quantity defined in Eq.~(\ref{eq:h0_qp}) turns out to be the e\-ner\-gy that can be attributed to the free ra\-dia\-tion as it is not interacting with matter.

The dissipative nature of the dy\-na\-mi\-cal sy\-stem~(\ref{eq:2lev_qp}) emerges as one sees that the following relationships
hold
\begin{equation}
\dot{H}_{0}=-4\alpha nH_{0},\ \dot{n}=kH_{0}\label{eq:h0.dot.n.dot.dissipative}
\end{equation}
as $\alpha$ and $k$ are positive constants, and as long as $n \geq 0$, this means that $\dot{H_0} \leq 0 $ and  $\dot{n} \geq 0$; all in all, the sy\-stem of Eqs.~(\ref{eq:h0.dot.n.dot.dissipative}), that are equivalent to Eqs.~(\ref{eq:2lev_qp}), simply describe the consumption of $H_0$ in favour of the quantity $n$. The sy\-stem~(\ref{eq:2lev_qp}) has an e\-ner\-gy dynamics similar to \textit{classical dissipation}, that points towards the formulation of it as a \textit{complete metriplectic sy\-stem}~\cite{Materassi2016}.

\section{\label{sec:second} Metriplectic Formulation}
In order to recognize a CMS equivalent to Eqs.~(\ref{eq:2lev_qp}), let us put those ODEs in the general form of a Hamiltonian sy\-stem ``perturbed'' by dissipative terms, the most general form of which reads:
\beq
\left\{\begin{array}{l}\dot{q}=\{q,H\}+\psi_q \\ \dot{p}=\{p,H\}+\psi_p \\ \dot{n}=\{n,H\}+\psi_n\end{array}\right. ,
\label{eq:2levels1}
\eeq
with $H(p,q,n)$ the \emph{total Hamiltonian}, $\{f,g\}=J^{ij}\partial_if\partial_jg$ the \emph{Poisson brackets} (PB) with
\beq
J^{ij}=\left(\begin{array}{ccc}0&1&0\\-1&0&0\\0&0&0\end{array}\right) , 
  \label{eq:J}
  \eeq
and improperly $i,j=q,p,n$.
Seeing that $\{n,A\}=0$ for any observable $A(q,p,n)$ is straightforward from Eq.~(\ref{eq:J}). This implies that any function $\mathcal{C}(n)$ is a Casimir; indeed, $\{\mathcal{C}(n),H_0\}=\mathcal{C}'(n)\{n,H_0\}=0$, so that one has:
\[
(\ref{eq:nndiss})\Rightarrow\dot{C}\left(n\right)=0.
\]

In order to express Eq.~(\ref{eq:2levels1}) as a metriplectic sy\-stem~\cite{Materassi2015,Materassi2016}, we define the \emph{metric brackets} $(f,g)=G^{ij}\partial_if\partial_jg$, constructing
\beq
G^{ij}=\left(\begin{array}{ccc}G^{qq}&G^{qp}&G^{qn}\\G^{qp}&G^{pp}&G^{pn}\\G^{qn}&G^{pn}&G^{nn}\end{array}\right)
\label{eq:G}
\eeq
as a symmetric, positive semi-definite matrix.

Eq.~(\ref{eq:2levels1}) will be put in the form of
\beq
\left\{\begin{array}{l}\dot{q}=\{q,H\}+\chi (q,S)\\ \dot{p}=\{p,H\}+\chi (p,S)\\ \dot{n}=\chi (n,S)\end{array}\right. ,
\label{eq:metriplectic1}
\eeq
where $\chi$ is a constant to be calculated once the desired $Z_\mathrm{eq}$ is defined. For the moment being, $\chi=\pm 1$ may be understood. Indeed, once defined the \emph{metriplectic (or Leibniz) brackets}
\beq
<<f,g>>:=\{f,g\}+(f,g),
\label{eq:metrb}
\eeq
the \emph{entropy} $S(q,p,n)$ and the \emph{free e\-ner\-gy} $F=H+\chi S$, if $\nabla S\in Ker(J)$ and $\nabla H\in Ker(G)$, namely,
\beq
J^{ij}\partial_jS=G^{ij}\partial_jH=0,
\label{eq:ker}
\eeq
then
\beq
\dot{g}=<<g,F>>=\{g,H\}+\chi(g,S).
\label{eq:dt}
\eeq

Eq.~(\ref{eq:ker}) implies that the entropy is a mere function of $n$.

\subsection{\label{ssec:first} The Metric Brackets Tensor}

Thanks to Eq.~(\ref{eq:ker}), the sy\-stem in Eq.~(\ref{eq:metriplectic1}) becomes
\beq
\left\{\begin{array}{l}\dot{q}=\partial_pH+\chi G^{qn}S'(n)\\ \dot{p}=\partial_qH+\chi G^{pn}S'(n)\\ \dot{n}=\chi G^{nn}S'(n)\end{array}\right. ,
\label{eq:metriplectic2}
\eeq
therefore our overriding concern is to determine the tensor $G$. In order to obtain such a result, we follow a linear algebraic procedure. Calculations are illustrated item by item in Appendix. The final result is:

\begin{widetext}
$$G_{\mathbb{E}}^{qq}=\frac{\partial_qH\partial_nH\left[b\partial_qH\partial_nH+2c\partial_pH\sqrt{(\partial_qH)^2+(\partial_pH)^2+(\partial_nH)^2}\right]}{\left[(\partial_qH)^2+(\partial_pH)^2\right]\left[(\partial_qH)^2+(\partial_pH)^2+(\partial_nH)^2\right]},$$
$$G_{\mathbb{E}}^{qp}=\frac{\partial_nH\left[b\partial_qH\partial_pH\partial_nH+c\left[(\partial_pH)^2-(\partial_qH)^2\right]\sqrt{(\partial_qH)^2+(\partial_pH)^2+(\partial_nH)^2}\right]}{\left[(\partial_qH)^2+(\partial_pH)^2\right]\left[(\partial_qH)^2+(\partial_pH)^2+(\partial_nH)^2\right]},$$
$$G_{\mathbb{E}}^{qn}=-\frac{b\partial_qH\partial_nH+c\partial_pH\sqrt{(\partial_qH)^2+(\partial_pH)^2+(\partial_nH)^2}}{(\partial_qH)^2+(\partial_pH)^2+(\partial_nH)^2},$$
$$G_{\mathbb{E}}^{pp}=\frac{b(\partial_pH)^2(\partial_nH)^2-2c\partial_qH\partial_pH\partial_nH\sqrt{(\partial_qH)^2+(\partial_pH)^2+(\partial_nH)^2}}{\left[(\partial_qH)^2+(\partial_pH)^2\right]\left[(\partial_qH)^2+(\partial_pH)^2+(\partial_nH)^2\right]},$$
$$G_{\mathbb{E}}^{pn}=\frac{-b\partial_pH\partial_nH+c\partial_qH\sqrt{(\partial_qH)^2+(\partial_pH)^2+(\partial_nH)^2}}{(\partial_qH)^2+(\partial_pH)^2+(\partial_nH)^2},$$
$$G_{\mathbb{E}}^{nn}=\frac{b\left[(\partial_qH)^2+(\partial_pH)^2\right]}{(\partial_qH)^2+(\partial_pH)^2+(\partial_nH)^2},$$
\end{widetext}
with
\beq
\left\{\begin{array}{l} b=\chi\frac{\psi_n}{S'(n)}\frac{(\partial_qH)^2+(\partial_pH)^2+(\partial_nH)^2}{(\partial_qH)^2+(\partial_pH)^2}\\ c=\chi\frac{\psi_p\partial_qH-\psi_q\partial_pH}{S'(n)}\frac{\sqrt{(\partial_qH)^2+(\partial_pH)^2+(\partial_nH)^2}}{(\partial_qH)^2+(\partial_pH)^2}\end{array}\right. .
\label{eq:bc}
\eeq

With reference to the Appendix, one attains a third equation from Eq.~(\ref{eq:G1}), namely,
\beq
\psi_q\partial_qH+\psi_p\partial_pH+\psi_n\partial_nH=0.
\label{eq:third}
\eeq
This last condition expresses the conservation of the Hamiltonian $H$ when the relationships~(\ref{eq:G1}) are enforced, that is precisely what is required by theory.

\subsection{\label{ssec:second} The Total Hamiltonian}
Considering Eq.~(\ref{eq:2lev_qp}), we fix
\beq
\left\{\begin{array}{l} \psi_q=-\alpha n q \\ \psi_p=-\alpha n p \\ \psi_n=\frac k8(q^2+p^2)^2 \end{array}\right. ,
\label{eq:psi}
\eeq
therefore $\partial_qH=\frac 12 q(q^2+p^2)$ and $\partial_pH=\frac 12 p(q^2+p^2)$, which imply that the total Hamiltonian reads: 
\beq
H(q,p,n)=H_0(q^2+p^2)+U(n),
\label{eq:totalH}
\eeq
being $H_0(q^2+p^2)$ the free ra\-dia\-tion Hamiltonian defined in Eq.~(\ref{eq:h0_qp}). In order to determine $U(n)$, we need to take into account Eq.~(\ref{eq:third}):
\beq
-\alpha n q\frac 12 q(q^2+p^2)-\alpha n p\frac 12 p(q^2+p^2)+\frac k8(q^2+p^2)^2U'(n)=0,
\label{eq:ODE_}
\eeq
whence
\beq
U(n)=\frac{2\alpha}k n^2+U_0.
\label{eq:U}
\eeq

\subsection{\label{ssec:third} Entropy and Equilibrium States}
We are now in the position of writing explicitely the free e\-ner\-gy $F$
$$F(q^2+p^2,n)=H_0(q^2+p^2)+U(n)+\chi S(n).$$
As the expression~(\ref{eq:U}) is used, one has
\begin{equation}
F\left(q^{2}+p^{2},n\right)=\frac{1}{8}\left(q^{2}+p^{2}\right)^{2}+\frac{2\alpha}{k}n^{2}+U_{0}+\chi S\left(n\right).\label{eq:explicit.F}
\end{equation}

E\-Qui\-Li\-Brium states of the sy\-stem must satisfy the condition
\beq
\delta F=\partial_qF\delta q+\partial_pF\delta p+\partial_nF\delta n=0,
\label{eq:eq_cond}
\eeq
that is:
\beq
q_{eq}=p_{eq}=0,\;\;S'(n)|_{n_{eq}}=-\chi\frac{4\alpha}k n_{eq}.
\label{eq:equilibrium}
\eeq
As expected, different entropy functions correspond to different e\-qui\-li\-brium points.

Some lines ago we anticipated that $\chi\rightarrow 0$ suppresses
the metric part of the dynamics: indeed here one sees that, in this
limit, the e\-qui\-li\-brium value of $n$ vanishes:
\[
\lim_{\chi\rightarrow0}n_{\mathrm{eq}}=0.
\]
This means putting oneself in the condition of an e\-qui\-li\-brium reached
without populating the atomic excited level (e.g., at $0{^\circ}K$
temperature), that does not mean turning off the matter-ra\-dia\-tion
coupling.

Before going to the conclusions, it is important to note that $\alpha$ and $k$ appear everywhere as a ratio: it would make sense to introduce an \textit{always finite}  constant $\kappa$ so that $k=\kappa \alpha$. This would reduce the non-dissipative condition~(\ref{eq:nndiss}) to the much simpler $\alpha \rightarrow 0$, that is, again, a statement about interactions, while $\chi \rightarrow 0$ would be a statement about the e\-qui\-li\-brium around which we are working.

\section{\label{sec:conclusions}Conclusions}
This work applies the metriplectic theory and Leibniz algebrae to a dissipative nonlinear optical phenomenon: the two-photon absorption by a two-level atom with ne\-gli\-gi\-ble spontaneous emission. Once the physical problem was formulated in terms of the conservative part $H_0$ of the total Hamiltonian $H$, the metric tensor $G$ and the metriplectic brackets $<<\cdot,\cdot>>$, we have found the mathematical expression of $H$ as function of the dy\-na\-mi\-cal va\-ria\-bles $q,p,n$. In particular, we have found the dissipative part $U$ of $H$, which depends only on the second level po\-pu\-la\-tion $n$. We have also found the free e\-ner\-gy $F$ and the e\-qui\-li\-brium states, varying with the de\-fi\-ni\-tion of entropy, as expected.

We believe that this manuscript opens the way to an ambitious research program in which the metriplectic formalism is used to explore irreversibility in nonlinear optics.
\section{Acknowledgments}
C.C. and G.M. acknowledge support from the Templeton Foundation (grant number 58277), the  H2020 QuantERA project QUOMPLEX (project ID 731473) and PRIN project NEMO (ref. 2015KEZNYM). They also acknowledge I. M. Deen for technical support with the computational resources.

\section*{Appendix}
Here we determine the tensor $G$. We start from Eq.~(\ref{eq:ker}) and look for an orthornormal basis $\mathbb{B}:=(\hat{n}_1,\hat{n}_2,\hat{n}_3)$, with $\hat{n}_3=\frac{\nabla H}{||\nabla H||}$, through a standard Gram-Schmidt process.
We find
\begin{widetext}
\beq
\hat{n}_1=\left(\begin{array}{c} -\frac{\partial_pH}{\sqrt{(\partial_qH)^2+(\partial_pH)^2}}\\ \frac{\partial_qH}{\sqrt{(\partial_qH)^2+(\partial_pH)^2}}\\ 0\end{array}\right),\;
  \hat{n}_2=\left(\begin{array}{c} -\frac{\partial_qH\partial_nH}{\sqrt{(\partial_qH)^2+(\partial_pH)^2}\sqrt{(\partial_qH)^2+(\partial_pH)^2+(\partial_nH)^2}}\\ -\frac{\partial_pH\partial_nH}{\sqrt{(\partial_qH)^2+(\partial_pH)^2}\sqrt{(\partial_qH)^2+(\partial_pH)^2+(\partial_nH)^2}}\\ \sqrt{\frac{(\partial_qH)^2+(\partial_pH)^2}{(\partial_qH)^2+(\partial_pH)^2+(\partial_nH)^2}}\end{array}\right),\;
  \hat{n}_3=\left(\begin{array}{c} \frac{\partial_qH}{\sqrt{(\partial_qH)^2+(\partial_pH)^2+(\partial_nH)^2}}\\ \frac{\partial_pH}{\sqrt{(\partial_qH)^2+(\partial_pH)^2+(\partial_nH)^2}}\\ \frac{\partial_nH}{\sqrt{(\partial_qH)^2+(\partial_pH)^2+(\partial_nH)^2}}\end{array}\right).
\label{eq:basisB}
\eeq
\end{widetext}

Then, we move from basis $\mathbb{B}$ to the canonical basis $\mathbb{E}=(\hat{e}_1,\hat{e}_2,\hat{e}_3)$, $\hat{e}_j=(\delta_{ij})^t_{i=1,2,3}$, by defining the unitary change of basis matrix
\begin{widetext}
\beq
C=\left(\begin{array}{ccc} -\frac{\partial_pH}{\sqrt{(\partial_qH)^2+(\partial_pH)^2}} & -\frac{\partial_qH\partial_nH}{\sqrt{(\partial_qH)^2+(\partial_pH)^2}\sqrt{(\partial_qH)^2+(\partial_pH)^2+(\partial_nH)^2}} & \frac{\partial_qH}{\sqrt{(\partial_qH)^2+(\partial_pH)^2+(\partial_nH)^2}}\\ \frac{\partial_qH}{\sqrt{(\partial_qH)^2+(\partial_pH)^2}} & -\frac{\partial_pH\partial_nH}{\sqrt{(\partial_qH)^2+(\partial_pH)^2}\sqrt{(\partial_qH)^2+(\partial_pH)^2+(\partial_nH)^2}} & \frac{\partial_pH}{\sqrt{(\partial_qH)^2+(\partial_pH)^2+(\partial_nH)^2}}\\ 0 & \sqrt{\frac{(\partial_qH)^2+(\partial_pH)^2}{(\partial_qH)^2+(\partial_pH)^2+(\partial_nH)^2}} & \frac{\partial_nH}{\sqrt{(\partial_qH)^2+(\partial_pH)^2+(\partial_nH)^2}}\end{array}\right),
\label{eq:change}
\eeq
\end{widetext}
whence
$$C^{-1}=C^t\mbox{ and }\mathbb{B}=\mathbb{E}C.$$
On $\mathbb{E}$, the tensor $G$ is expressed in Eq.~(\ref{eq:G}), but, in order to obey Eq.~(\ref{eq:ker}), on $\mathbb{B}$ it must be $\nabla H$-transverse, that is,

\beq
G_{\mathbb{B}}=\left(\begin{array}{ccc} a& c& 0\\c&b&0\\0&0&0\end{array}\right),
\label{eq:GB}
\eeq
with $a,b,c\in\mathbb{R}$ and $ab-c^2\neq0$.
Since $G_{\mathbb{E}}=CG_{\mathbb{B}}C^t$, it turns out that
\begin{widetext}
$$G_{\mathbb{E}}^{qq}=\frac{a(\partial_pH)^2\left[(\partial_qH)^2+(\partial_pH)^2+(\partial_nH)^2\right]+\partial_qH\partial_nH\left[b\partial_qH\partial_nH+2c\partial_pH\sqrt{(\partial_qH)^2+(\partial_pH)^2+(\partial_nH)^2}\right]}{\left[(\partial_qH)^2+(\partial_pH)^2\right]\left[(\partial_qH)^2+(\partial_pH)^2+(\partial_nH)^2\right]},$$
$$G_{\mathbb{E}}^{qp}=\frac{-a\partial_qH\partial_pH\left[(\partial_qH)^2+(\partial_pH)^2+(\partial_nH)^2\right]+\partial_nH\left[b\partial_qH\partial_pH\partial_nH+c\left[(\partial_pH)^2-(\partial_qH)^2\right]\sqrt{(\partial_qH)^2+(\partial_pH)^2+(\partial_nH)^2}\right]}{\left[(\partial_qH)^2+(\partial_pH)^2\right]\left[(\partial_qH)^2+(\partial_pH)^2+(\partial_nH)^2\right]},$$
$$G_{\mathbb{E}}^{qn}=-\frac{b\partial_qH\partial_nH+c\partial_pH\sqrt{(\partial_qH)^2+(\partial_pH)^2+(\partial_nH)^2}}{(\partial_qH)^2+(\partial_pH)^2+(\partial_nH)^2},$$
$$G_{\mathbb{E}}^{pp}=\frac{b(\partial_pH)^2(\partial_nH)^2+\partial_qH\left\{-2c\partial_pH\partial_nH\sqrt{(\partial_qH)^2+(\partial_pH)^2+(\partial_nH)^2}+a\partial_qH\left[(\partial_qH)^2+(\partial_pH)^2+(\partial_nH)^2\right]\right\}}{\left[(\partial_qH)^2+(\partial_pH)^2\right]\left[(\partial_qH)^2+(\partial_pH)^2+(\partial_nH)^2\right]},$$
$$G_{\mathbb{E}}^{pn}=\frac{-b\partial_pH\partial_nH+c\partial_qH\sqrt{(\partial_qH)^2+(\partial_pH)^2+(\partial_nH)^2}}{(\partial_qH)^2+(\partial_pH)^2+(\partial_nH)^2},$$
$$G_{\mathbb{E}}^{nn}=\frac{b\left[(\partial_qH)^2+(\partial_pH)^2\right]}{(\partial_qH)^2+(\partial_pH)^2+(\partial_nH)^2}.$$
\end{widetext}
By comparing Eq.~(\ref{eq:2levels1}) and Eq.~(\ref{eq:metriplectic2}), we get
\beq
\left\{\begin{array}{l} G^{qn}=\chi\frac{\psi_q}{S'(n)}\\ G^{pn}=\chi\frac{\psi_p}{S'(n)}\\ G^{nn}=\chi\frac{\psi_q}{S'(n)}\end{array}\right.,
\label{eq:G1}
\eeq
which is indepedent of $a$: hence we may fix $a=0$ without loss of generality.
Finally, we solve Eq.~(\ref{eq:G1}) for the parameter $b,c$, and obtain Eq.~(\ref{eq:bc}).


\begin{thebibliography}{16}%
\makeatletter
\providecommand \@ifxundefined [1]{%
 \@ifx{#1\undefined}
}%
\providecommand \@ifnum [1]{%
 \ifnum #1\expandafter \@firstoftwo
 \else \expandafter \@secondoftwo
 \fi
}%
\providecommand \@ifx [1]{%
 \ifx #1\expandafter \@firstoftwo
 \else \expandafter \@secondoftwo
 \fi
}%
\providecommand \natexlab [1]{#1}%
\providecommand \enquote  [1]{``#1''}%
\providecommand \bibnamefont  [1]{#1}%
\providecommand \bibfnamefont [1]{#1}%
\providecommand \citenamefont [1]{#1}%
\providecommand \href@noop [0]{\@secondoftwo}%
\providecommand \href [0]{\begingroup \@sanitize@url \@href}%
\providecommand \@href[1]{\@@startlink{#1}\@@href}%
\providecommand \@@href[1]{\endgroup#1\@@endlink}%
\providecommand \@sanitize@url [0]{\catcode `\\12\catcode `\$12\catcode
  `\&12\catcode `\#12\catcode `\^12\catcode `\_12\catcode `\%12\relax}%
\providecommand \@@startlink[1]{}%
\providecommand \@@endlink[0]{}%
\providecommand \url  [0]{\begingroup\@sanitize@url \@url }%
\providecommand \@url [1]{\endgroup\@href {#1}{\urlprefix }}%
\providecommand \urlprefix  [0]{URL }%
\providecommand \Eprint [0]{\href }%
\providecommand \doibase [0]{http://dx.doi.org/}%
\providecommand \selectlanguage [0]{\@gobble}%
\providecommand \bibinfo  [0]{\@secondoftwo}%
\providecommand \bibfield  [0]{\@secondoftwo}%
\providecommand \translation [1]{[#1]}%
\providecommand \BibitemOpen [0]{}%
\providecommand \bibitemStop [0]{}%
\providecommand \bibitemNoStop [0]{.\EOS\space}%
\providecommand \EOS [0]{\spacefactor3000\relax}%
\providecommand \BibitemShut  [1]{\csname bibitem#1\endcsname}%
\let\auto@bib@innerbib\@empty
\bibitem [{\citenamefont {Marcucci}\ and\ \citenamefont
  {Conti}(2016)}]{Marcucci2016}%
  \BibitemOpen
  \bibfield  {author} {\bibinfo {author} {\bibfnamefont {G.}~\bibnamefont
  {Marcucci}}\ and\ \bibinfo {author} {\bibfnamefont {C.}~\bibnamefont
  {Conti}},\ }\href@noop {} {\bibfield  {journal} {\bibinfo  {journal} {Phys.
  Rev. A}\ }\textbf {\bibinfo {volume} {94}},\ \bibinfo {pages} {052136}
  (\bibinfo {year} {2016})}\BibitemShut {NoStop}%
\bibitem [{\citenamefont {Garmon}\ \emph {et~al.}(2012)\citenamefont {Garmon},
  \citenamefont {Petrosky}, \citenamefont {Simine},\ and\ \citenamefont
  {Segal}}]{Garmon2012}%
  \BibitemOpen
  \bibfield  {author} {\bibinfo {author} {\bibfnamefont {S.}~\bibnamefont
  {Garmon}}, \bibinfo {author} {\bibfnamefont {T.}~\bibnamefont {Petrosky}},
  \bibinfo {author} {\bibfnamefont {L.}~\bibnamefont {Simine}}, \ and\ \bibinfo
  {author} {\bibfnamefont {D.}~\bibnamefont {Segal}},\ }\href@noop {}
  {\bibfield  {journal} {\bibinfo  {journal} {Fortschritte der Physik}\
  }\textbf {\bibinfo {volume} {61}},\ \bibinfo {pages} {261} (\bibinfo {year}
  {2012})}\BibitemShut {NoStop}%
\bibitem [{\citenamefont {Turski}(1987)}]{Turski1987}%
  \BibitemOpen
  \bibfield  {author} {\bibinfo {author} {\bibfnamefont {L.~A.}\ \bibnamefont
  {Turski}},\ }\href@noop {} {\bibfield  {journal} {\bibinfo  {journal}
  {Physics Letters A}\ }\textbf {\bibinfo {volume} {125}},\ \bibinfo {pages}
  {461} (\bibinfo {year} {1987})}\BibitemShut {NoStop}%
\bibitem [{\citenamefont {Turski}(1996)}]{Turski1996}%
  \BibitemOpen
  \bibfield  {author} {\bibinfo {author} {\bibfnamefont {L.~A.}\ \bibnamefont
  {Turski}},\ }in\ \href@noop {} {\emph {\bibinfo {booktitle} {From Quantum
  Mechanics to Technology}}},\ \bibinfo {editor} {edited by\ \bibinfo {editor}
  {\bibfnamefont {Z.}~\bibnamefont {Petru}}, \bibinfo {editor} {\bibfnamefont
  {J.}~\bibnamefont {Przystawa}}, \ and\ \bibinfo {editor} {\bibfnamefont
  {K.}~\bibnamefont {Rapcewicz}}}\ (\bibinfo  {publisher} {Springer Berlin
  Heidelberg},\ \bibinfo {address} {Berlin, Heidelberg},\ \bibinfo {year}
  {1996})\BibitemShut {NoStop}%
\bibitem [{\citenamefont {Materassi}\ and\ \citenamefont
  {Tassi}(2012{\natexlab{a}})}]{Materassi2012bis}%
  \BibitemOpen
  \bibfield  {author} {\bibinfo {author} {\bibfnamefont {M.}~\bibnamefont
  {Materassi}}\ and\ \bibinfo {author} {\bibfnamefont {E.}~\bibnamefont
  {Tassi}},\ }\href@noop {} {\bibfield  {journal} {\bibinfo  {journal}
  {Intellectual Archive}\ }\textbf {\bibinfo {volume} {1}},\ \bibinfo {pages}
  {45} (\bibinfo {year} {2012}{\natexlab{a}})}\BibitemShut {NoStop}%
\bibitem [{\citenamefont {Morrison}(1984)}]{Morrison1984}%
  \BibitemOpen
  \bibfield  {author} {\bibinfo {author} {\bibfnamefont {P.~J.}\ \bibnamefont
  {Morrison}},\ }\href@noop {} {\bibfield  {journal} {\bibinfo  {journal}
  {Physics Letters A}\ }\textbf {\bibinfo {volume} {100}},\ \bibinfo {pages}
  {423} (\bibinfo {year} {1984})}\BibitemShut {NoStop}%
\bibitem [{\citenamefont {Materassi}\ and\ \citenamefont
  {Tassi}(2012{\natexlab{b}})}]{Materassi2012}%
  \BibitemOpen
  \bibfield  {author} {\bibinfo {author} {\bibfnamefont {M.}~\bibnamefont
  {Materassi}}\ and\ \bibinfo {author} {\bibfnamefont {E.}~\bibnamefont
  {Tassi}},\ }\href@noop {} {\bibfield  {journal} {\bibinfo  {journal} {Physica
  D}\ }\textbf {\bibinfo {volume} {241}},\ \bibinfo {pages} {729} (\bibinfo
  {year} {2012}{\natexlab{b}})}\BibitemShut {NoStop}%
\bibitem [{\citenamefont {Morrison}(1986)}]{Morrison1986}%
  \BibitemOpen
  \bibfield  {author} {\bibinfo {author} {\bibfnamefont {P.~J.}\ \bibnamefont
  {Morrison}},\ }\href@noop {} {\bibfield  {journal} {\bibinfo  {journal}
  {Physica D}\ }\textbf {\bibinfo {volume} {18}},\ \bibinfo {pages} {410}
  (\bibinfo {year} {1986})}\BibitemShut {NoStop}%
\bibitem [{\citenamefont {Guha}(2007)}]{Guha2007}%
  \BibitemOpen
  \bibfield  {author} {\bibinfo {author} {\bibfnamefont {P.}~\bibnamefont
  {Guha}},\ }\href@noop {} {\bibfield  {journal} {\bibinfo  {journal} {Journal
  of Mathematical Analysis and Applications}\ }\textbf {\bibinfo {volume}
  {326}},\ \bibinfo {pages} {121} (\bibinfo {year} {2007})}\BibitemShut
  {NoStop}%
\bibitem [{\citenamefont {Materassi}(2016)}]{Materassi2016}%
  \BibitemOpen
  \bibfield  {author} {\bibinfo {author} {\bibfnamefont {M.}~\bibnamefont
  {Materassi}},\ }\href@noop {} {\bibfield  {journal} {\bibinfo  {journal}
  {Entropy}\ }\textbf {\bibinfo {volume} {18}},\ \bibinfo {pages} {304}
  (\bibinfo {year} {2016})}\BibitemShut {NoStop}%
\bibitem [{\citenamefont {Boyd}(2008)}]{Boyd2008}%
  \BibitemOpen
  \bibfield  {author} {\bibinfo {author} {\bibfnamefont {R.~W.}\ \bibnamefont
  {Boyd}},\ }\href@noop {} {\emph {\bibinfo {title} {Nonlinear Optics}}},\
  \bibinfo {edition} {3rd}\ ed.\ (\bibinfo  {publisher} {Academic Press},\
  \bibinfo {address} {Burlington},\ \bibinfo {year} {2008})\BibitemShut
  {NoStop}%
\bibitem [{\citenamefont {Moloney}\ and\ \citenamefont
  {Newell}(2004)}]{Moloney2004}%
  \BibitemOpen
  \bibfield  {author} {\bibinfo {author} {\bibfnamefont {J.}~\bibnamefont
  {Moloney}}\ and\ \bibinfo {author} {\bibfnamefont {A.}~\bibnamefont
  {Newell}},\ }\href@noop {} {\emph {\bibinfo {title} {Nonlinear Optics}}},\
  Advanced Book Program\ (\bibinfo  {publisher} {Avalon Publishing},\ \bibinfo
  {year} {2004})\BibitemShut {NoStop}%
\bibitem [{\citenamefont {Feng}\ \emph {et~al.}(1997)\citenamefont {Feng},
  \citenamefont {Moloney}, \citenamefont {Newell}, \citenamefont {Wright},
  \citenamefont {Cook}, \citenamefont {Kennedy}, \citenamefont {Hammer},
  \citenamefont {Rockwell},\ and\ \citenamefont {Thompson}}]{Feng1997}%
  \BibitemOpen
  \bibfield  {author} {\bibinfo {author} {\bibfnamefont {Q.}~\bibnamefont
  {Feng}}, \bibinfo {author} {\bibfnamefont {J.~V.}\ \bibnamefont {Moloney}},
  \bibinfo {author} {\bibfnamefont {A.~C.}\ \bibnamefont {Newell}}, \bibinfo
  {author} {\bibfnamefont {E.~M.}\ \bibnamefont {Wright}}, \bibinfo {author}
  {\bibfnamefont {K.}~\bibnamefont {Cook}}, \bibinfo {author} {\bibfnamefont
  {P.~K.}\ \bibnamefont {Kennedy}}, \bibinfo {author} {\bibfnamefont {D.~X.}\
  \bibnamefont {Hammer}}, \bibinfo {author} {\bibfnamefont {B.~A.}\
  \bibnamefont {Rockwell}}, \ and\ \bibinfo {author} {\bibfnamefont {C.~R.}\
  \bibnamefont {Thompson}},\ }\href@noop {} {\bibfield  {journal} {\bibinfo
  {journal} {IEEE Journal of Quantum Electronics}\ }\textbf {\bibinfo {volume}
  {33}},\ \bibinfo {pages} {127} (\bibinfo {year} {1997})}\BibitemShut
  {NoStop}%
\bibitem [{\citenamefont {Mlejnek}\ \emph {et~al.}(1998)\citenamefont
  {Mlejnek}, \citenamefont {Wright},\ and\ \citenamefont
  {Moloney}}]{Mlejnek1998}%
  \BibitemOpen
  \bibfield  {author} {\bibinfo {author} {\bibfnamefont {M.}~\bibnamefont
  {Mlejnek}}, \bibinfo {author} {\bibfnamefont {E.~M.}\ \bibnamefont {Wright}},
  \ and\ \bibinfo {author} {\bibfnamefont {J.~V.}\ \bibnamefont {Moloney}},\
  }\href@noop {} {\bibfield  {journal} {\bibinfo  {journal} {Opt. Lett.}\
  }\textbf {\bibinfo {volume} {23}},\ \bibinfo {pages} {382} (\bibinfo {year}
  {1998})}\BibitemShut {NoStop}%
\bibitem [{\citenamefont {Schwarz}\ \emph {et~al.}(2000)\citenamefont
  {Schwarz}, \citenamefont {Rambo}, \citenamefont {Diels}, \citenamefont
  {Kolesik}, \citenamefont {Wright},\ and\ \citenamefont
  {Moloney}}]{Schwarz2000}%
  \BibitemOpen
  \bibfield  {author} {\bibinfo {author} {\bibfnamefont {J.}~\bibnamefont
  {Schwarz}}, \bibinfo {author} {\bibfnamefont {P.}~\bibnamefont {Rambo}},
  \bibinfo {author} {\bibfnamefont {J.-C.}\ \bibnamefont {Diels}}, \bibinfo
  {author} {\bibfnamefont {M.}~\bibnamefont {Kolesik}}, \bibinfo {author}
  {\bibfnamefont {E.~M.}\ \bibnamefont {Wright}}, \ and\ \bibinfo {author}
  {\bibfnamefont {J.~V.}\ \bibnamefont {Moloney}},\ }\href@noop {} {\bibfield
  {journal} {\bibinfo  {journal} {Optics Communications}\ }\textbf {\bibinfo
  {volume} {180}},\ \bibinfo {pages} {383 } (\bibinfo {year}
  {2000})}\BibitemShut {NoStop}%
\bibitem [{\citenamefont {Materassi}(2015)}]{Materassi2015}%
  \BibitemOpen
  \bibfield  {author} {\bibinfo {author} {\bibfnamefont {M.}~\bibnamefont
  {Materassi}},\ }\href@noop {} {\bibfield  {journal} {\bibinfo  {journal}
  {Entropy}\ }\textbf {\bibinfo {volume} {17}},\ \bibinfo {pages} {1329}
  (\bibinfo {year} {2015})}\BibitemShut {NoStop}%
\end{thebibliography}

%

\end{document}